\input harvmac.tex

\input epsf.tex

\def\figin{\epsfcheck\figin}\def\figins{\epsfcheck\figins}
\def\epsfcheck{\ifx\epsfbox\UnDeFiNeD
\message{(NO epsf.tex, FIGURES WILL BE IGNORED)}
\gdef\figin##1{\vskip2in}\gdef\figins##1{\hskip.5in}
\else\message{(FIGURES WILL BE INCLUDED)}%
\gdef\figin##1{##1}\gdef\figins##1{##1}\fi}
\def\DefWarn#1{}
\def\figinsert{\goodbreak\midinsert}
\def\ifig#1#2#3{\DefWarn#1\xdef#1{fig.~\the\figno}
\writedef{#1\leftbracket fig.\noexpand~\the\figno}
\figinsert\figin{\centerline{#3}}\medskip\centerline{\vbox{\baselineskip12pt
\advance\hsize by -1truein\noindent\footnotefont{\bf
Fig.~\the\figno:} #2}}
\bigskip\endinsert\global\advance\figno by1}



\lref\chang{ A. Chang, J. Qing and P. Yang,
arXiv:math/0512376
}

\lref\Bender{
  C.~M.~Bender, P.~D.~Mannheim,
Phys.\ Rev.\ Lett.\  {\bf 100}, 110402 (2008).
[arXiv:0706.0207 [hep-th]].
}

\lref\anderson{
M. Anderson,
Math. Res. Lett., 8 (2001) 171-188, arXiv:math/0011051
}

\lref\MannheimDS{
  P.~D.~Mannheim,
[arXiv:1101.2186 [hep-th]].
}

\lref\Adler{
  S.~L.~Adler,
Rev.\ Mod.\ Phys.\  {\bf 54}, 729 (1982).
}

\lref\LiuBU{
  H.~Liu, A.~A.~Tseytlin,
Nucl.\ Phys.\  {\bf B533}, 88-108 (1998).
[hep-th/9804083].
}

\lref\BalasubramanianPQ{
  V.~Balasubramanian, E.~G.~Gimon, D.~Minic, J.~Rahmfeld,
Phys.\ Rev.\  {\bf D63}, 104009 (2001).
[hep-th/0007211].
}

\lref\BergshoeffIS{
  E.~Bergshoeff, M.~de Roo, B.~de Wit,
Nucl.\ Phys.\  {\bf B182}, 173 (1981).
  B.~de Wit, J.~W.~van Holten, A.~Van Proeyen,
Nucl.\ Phys.\  {\bf B184}, 77 (1981).
}

\lref\BerkovitsWitten{
  N.~Berkovits, E.~Witten,
JHEP {\bf 0408}, 009 (2004).
[hep-th/0406051].
}

\lref\Fradkin{
  E.~S.~Fradkin, A.~A.~Tseytlin,
Phys.\ Rept.\  {\bf 119}, 233-362 (1985).
}

\lref\maldaInfl{
  J.~M.~Maldacena,
JHEP {\bf 0305}, 013 (2003).
[astro-ph/0210603].
}

\lref\StromingerConformal{
  A.~Strominger, V.~P.~Nair,
Phys.\ Rev.\  {\bf D30}, 2528 (1984).
}
\lref\Instability{
  D.~G.~Boulware, G.~T.~Horowitz, A.~Strominger,
Phys.\ Rev.\ Lett.\  {\bf 50}, 1726 (1983).
}

\lref\Harlow{
  D.~Harlow, D.~Stanford,
[arXiv:1104.2621 [hep-th]].
}
\lref\tHooft{
  G.~'t Hooft,
  arXiv:1104.4543 [gr-qc].
}
\lref\WaldNT{
  R.~M.~Wald,
  Phys.\ Rev.\  D {\bf 48}, 3427 (1993)
  [arXiv:gr-qc/9307038].
}

\lref\JacobsonVJ{
  T.~Jacobson, G.~Kang, R.~C.~Myers,
Phys.\ Rev.\  {\bf D49}, 6587-6598 (1994).
[gr-qc/9312023].
}

\lref\Starobinsky{
A. A. Starobinsky,
JETP Lett. 37, 66 (1983).
}

\lref\Emparan{
  R.~Emparan,
JHEP {\bf 9906}, 036 (1999).
[hep-th/9906040].
}

\lref\FG{
 C. Fefferman and C.R. Graham, �Conformal Invariants�, in Elie Cartan et les
Math�ematiques d�aujourd�hui (Ast�erisque, 1985) 95.
}

\lref\Axial{
  H.~Romer, P.~van Nieuwenhuizen,
Phys.\ Lett.\  {\bf B162}, 290 (1985).
}

\lref\FradkinBeta{
  E.~S.~Fradkin, A.~A.~Tseytlin,
Phys.\ Lett.\  {\bf B134}, 187 (1984).
}

\lref\LiuBU{
  H.~Liu, A.~A.~Tseytlin,
Nucl.\ Phys.\  {\bf B533}, 88-108 (1998).
[hep-th/9804083].
}

\lref\Henningson{
  M.~Henningson, K.~Skenderis,
JHEP {\bf 9807}, 023 (1998).
[hep-th/9806087].
}

\lref\changToAppear{
A. Chang, S-Y. S. Chen, P. Yang, Private Communication.
 }

\lref\Pope{   H.~Lu, C.~N.~Pope,
Phys.\ Rev.\ Lett.\  {\bf 106}, 181302 (2011).
[arXiv:1101.1971 [hep-th]].
  S.~Deser, H.~Liu, H.~Lu, C.~N.~Pope, T.~C.~Sisman, B.~Tekin,
Phys.\ Rev.\  {\bf D83}, 061502 (2011).
[arXiv:1101.4009 [hep-th]].
}

\lref\HP{
  S.~W.~Hawking, D.~N.~Page,
Commun.\ Math.\ Phys.\  {\bf 87}, 577 (1983).
}

\lref\Marino{
  N.~Drukker, M.~Marino, P.~Putrov,
[arXiv:1007.3837 [hep-th]].
}
\lref\Kapustin{
  A.~Kapustin, B.~Willett, I.~Yaakov,
JHEP {\bf 1010}, 013 (2010).
[arXiv:1003.5694 [hep-th]].
}
\lref\Jafferis{
  D.~L.~Jafferis,
[arXiv:1012.3210 [hep-th]].
}

\lref\Wald{
  R.~M.~Wald,
  ``General Relativity,''
Chicago, Usa: Univ. Pr. ( 1984) 491p.
}

\lref\HH{
  J.~B.~Hartle, S.~W.~Hawking,
Phys.\ Rev.\  {\bf D28}, 2960-2975 (1983).
}

\lref\LarsenPF{
  F.~Larsen and R.~McNees,
  JHEP {\bf 0307}, 051 (2003)
  [arXiv:hep-th/0307026].
}
\lref\McFaddenFG{
  P.~McFadden, K.~Skenderis,
Phys.\ Rev.\  {\bf D81}, 021301 (2010).
[arXiv:0907.5542 [hep-th]].
}

\lref\RiegertZZ{
  R.~J.~Riegert,
Phys.\ Rev.\ Lett.\  {\bf 53}, 315-318 (1984).
}
\lref\DeserUP{
  S.~Deser, B.~Tekin,
Class.\ Quant.\ Grav.\  {\bf 20}, 4877-4884 (2003).
[gr-qc/0306114].
}
\lref\Bach{ R. Bach, Math. Zeit., {\bf 9 } , 110 (1921). }

\lref\RiegertHF{
  R.~J.~Riegert,
Phys.\ Lett.\  {\bf A105}, 110-112 (1984).
}
\lref\SmilgaPR{
  A.~V.~Smilga,
SIGMA {\bf 5}, 017 (2009).
[arXiv:0808.0139 [quant-ph]].
}

\lref\Oleath{
  O.~Miskovic, R.~Olea,
Phys.\ Rev.\  {\bf D79}, 124020 (2009).
[arXiv:0902.2082 [hep-th]].
}

\lref\Oleaf{
   R.~Aros, M.~Contreras, R.~Olea, R.~Troncoso, J.~Zanelli,
Phys.\ Rev.\ Lett.\  {\bf 84}, 1647-1650 (2000).
[gr-qc/9909015].
}
\lref\Oleas{
 R.~Olea,
JHEP {\bf 0506}, 023 (2005).
[hep-th/0504233].
}

\lref\Skenderisf{
  S.~de Haro, S.~N.~Solodukhin and K.~Skenderis,
  Commun.\ Math.\ Phys.\  {\bf 217}, 595 (2001)
  [arXiv:hep-th/0002230].
}

\lref\Skenderiss{
  K.~Skenderis,
  Class.\ Quant.\ Grav.\  {\bf 19}, 5849 (2002)
  [arXiv:hep-th/0209067].
}
\lref\BKraus{
  V.~Balasubramanian and P.~Kraus,
  Commun.\ Math.\ Phys.\  {\bf 208}, 413 (1999)
  [arXiv:hep-th/9902121].
}
\lref\MetsaevFQ{
  R.~R.~Metsaev,
  arXiv:0707.4437 [hep-th].
}

\lref\DeserPartial{
  S.~Deser, R.~I.~Nepomechie,
Annals Phys.\  {\bf 154}, 396 (1984).
  S.~Deser and A.~Waldron,
  Phys.\ Rev.\ Lett.\  {\bf 87}, 031601 (2001)
  [arXiv:hep-th/0102166].
}
\Title{\vbox{\baselineskip12pt \hbox{} \hbox{
} }} {\vbox{\centerline{  Einstein Gravity from Conformal Gravity
  }
\centerline{
  }
}}
\bigskip
\centerline{  Juan Maldacena }
\bigskip
\centerline{ \it  School of Natural Sciences, Institute for
Advanced Study} \centerline{\it Princeton, NJ 08540, USA}

\vskip .3in \noindent

We show that that four dimensional conformal gravity plus a simple Neumann boundary condition can be used
to get the semiclassical (or tree level)
wavefunction of the universe of four dimensional asymptotically de-Sitter or Euclidean anti-de Sitter
spacetimes.
This  simple Neumann  boundary condition selects
the Einstein solution  out of the more numerous solutions of conformal gravity.
It thus  removes the ghosts of conformal gravity from this computation.

In the case of a five dimensional pure gravity theory with a positive cosmological constant
we show that the late time superhorizon tree level probability measure, $|\Psi [ g ]|^2$,
for its four dimensional spatial slices is given by the action of Euclidean
four dimensional conformal gravity.


 \Date{ }


\newsec{Introduction}

Conformal gravity is an intriguing theory of gravity. It is a theory of gravity in four
dimensions with  an action given by
the square of the Weyl tensor, $S_{\rm conf} = \int d^4 x \sqrt{g} \,  W^2$. It is a theory sensitive to angles, but not
  distances. In fact, a Weyl transformation of the metric, $g_{\mu \nu} \to \Omega^2(x) g_{\mu \nu}$,
is an exact symmetry of this action.
It has appeared periodically in the literature for various reasons. It was considered
as a possible UV completion of gravity \refs{\Adler,\MannheimDS,\tHooft}, and references therein.
 It was also useful for constructing supergravity theories, see e.g.  \BergshoeffIS .
It has recently emerged  from the twistor string theory
 \BerkovitsWitten . It has also appeared as a counter term in
$AdS_5$ or $CFT_4$ computations \refs{\LiuBU,\BalasubramanianPQ}.
It was seldom taken seriously because it has ghosts, due to the
fact that the equations of motion are fourth order\foot{See \Bender\ for ideas on how to deal with
ghosts and \SmilgaPR\ for some criticisms of that idea.
}.

Here we would like to point out a couple of  interesting  connections between
conformal gravity and ordinary gravity in $AdS$ or $dS$.
We show that
 conformal gravity with certain future boundary conditions is equivalent
to ordinary gravity in asymptotically de-Sitter space. Alternatively, we can say that by setting the ghost fields to zero in the future of de-Sitter,
we get a wavefunctional  for the metric which is the same as the
 one given by  Bunch Davies (or Hartle Hawking) at tree level.
 A similar relation is present for  Euclidean  spaces which are asymptotically $EAdS$, or hyperbolic
space.

It was observed in \anderson\  that the renormalized on shell
action of four dimensional Einstein gravity in asymptotically
hyperbolic Einstein spaces
is given by the the action of conformal gravity. As usual, this action is
 evaluated on a solution of Einstein gravity.
It is a well known fact, that the solutions of Einstein gravity are also solutions of conformal gravity.
But conformal gravity has other solutions. If we were able to select, in a simple way, the solutions
of Einstein gravity from the solutions of conformal gravity, then we can forget about the Einstein
action and use  instead the conformal gravity action in the bulk. Actually,
 it is very easy to select the solutions of Einstein gravity.  We simply need to
impose a Neumann boundary condition on the metric at the boundary. Then this simple boundary condition is
eliminating the ghosts and rendering the theory equivalent to ordinary pure
Einstein gravity with a cosmological constant.

The discussion in this paper is purely classical, or tree level, but it is non-linear. And it would be
interesting if one could somehow use this to construct a full quantum theory based on conformal
gravity alone, since an  ${\cal N}=4$ supersymmetric version of conformal gravity
is possibly finite, see  \Fradkin\ for a review.

It was  observed in \refs{\LiuBU,\Henningson}  that the action of four dimensional
conformal gravity appears as a logarithmically divergent counterterm in   computations in
{\rm five} dimensional asymptotically Hyperbolic spaces. It appears as the coefficient of
the holographic Weyl anomaly. An interesting situation arises when we consider the de-Sitter
version of this computation.
In de-Sitter space, one is often interested in computing the probability measure for the
spatial slices at superhorizon distances $\left|\Psi[g]\right|^2 $. This is the tree level solution of
the (superhorizon)
measure problem in such a universe. It turns out that this probability measure is given by the
action of conformal gravity in four dimensions $ |\Psi[g]|^2 = e^{ - S_{\rm conf} [g] } $.
This is due to two facts. First the fact noted in \refs{\maldaInfl,\McFaddenFG,\Harlow}   that the
de-Sitter wavefunction can be computed by a certain analytic continuation from the Euclidean AdS one.
The only term that becomes real is a term that comes from analytically continuing the logarithmic
divergence. This produces a finite term given by  the action of conformal gravity.

This paper is organized as follows. In section two we discuss a conformally
coupled scalar field with a fourth derivative action. This serves as a toy example for conformal
gravity. In section three we discuss the relation between conformal gravity with a boundary
condition and ordinary Einstein gravity in $AdS$ or $dS$. In section four we consider the black hole
contributions to the partition function of conformal gravity. In section five we make a side
comment regarding the Hartle Hawking measure factor and the 3-sphere partition function of a possible
dual boundary $CFT$. In section six we argue that the probability measure of five dimensional
de Sitter gravity is given by the action of four dimensional conformal gravity. We end with
a discussion.

\newsec{Conformal scalar field }

Before we discuss conformal gravity it is convenient to discuss the simpler case of
a conformally coupled field with a fourth derivative action.

The reader is probably familiar with the fact that the action
\eqn\dimones{
S = { 1 \over 2} \int d^4 x \sqrt{g} \left[  ( \nabla \phi)^2 + { 1 \over 6 } \phi^2 R \right]
 }
  describes a conformally coupled field with dimension one. The action is invariant under
  $g \to \Omega^2 g$,
  $\phi \to \Omega^{-1} \phi$.

Here we are more interested in considering a conformally coupled field, $C$, of dimension zero.
This is more similar to what we have for the metric, which also has dimension zero.
The action is then of fourth order
\eqn\actfour{
S = {1 \over 2 } \int d^4 x \sqrt{g} \left[   ( \nabla^2 C )^2  -2 ( R_{\mu \nu} - { 1 \over 3 } g_{\mu \nu} R )
\partial_\mu   C \partial_\nu C \right]
}
The curvature couplings are necessary for Weyl invariance ( $g \to \Omega^2 g$,
  $C \to C$)
\Fradkin , and are  analogous
 to the usual one in \dimones .

Around flat space we simply have the fourth order equation $ ( \partial^2)^2 C =0$ and
the  solutions are easy to find.  If $t$ is a time coordinate, then
for a given spatial momentum $\vec k$
the four solutions are $C = e^{ \pm i |\vec k | t }, ~ t e^{ \pm i |\vec k | t }$.
As emphasized in \BerkovitsWitten , the Hamiltonian is not diagonalizable, and it has a Jordan form. This is due
to $t$ factor in the second solution.   If we choose $t$ as Euclidean time, we also have similar looking solutions
but with $e^{ \pm |k| t}$ instead.

We can now consider the same problem in $AdS_4$ with the metric
\eqn\metr{
 ds^2 = {   dz^2 + dx^2 \over z^2 }
 }
 This metric is equivalent, up to a Weyl transformation, to the flat space metric. Since this
 field is conformaly coupled we expect that the answers are the same as the ones we would obtain in
 flat space.
 It is interesting, nevertheless, to consider the field action in $AdS$ space. In that case \actfour\
 simplifies and gives
 \eqn\actf{
  S = - { 1 \over 2 } \int_{AdS_4} \sqrt{g} \left[  ( \nabla^2  C)^2 - 2 (\nabla C)^2 \right]
  }
  Introducing an extra field this can be rewritten as
  \eqn\actfi{ S = \int_{AdS_4} \sqrt{g} \left\{   [ \nabla (C + \varphi )]^2 - [ (\nabla \varphi)^2 - 2 \varphi^2 ]  \right\}
  }
  Integrating out $\varphi$ we get back to \actf , see \MetsaevFQ . The equation of motion for $\varphi$ sets it equal to
  $\varphi = { 1 \over 2 }  \nabla^2 C$.  Defining $\tilde C = C + \varphi$, we see that
  we have two scalar fields, one massless and the other with $m^2 = - 2$, which is a tachyon in
   the allowed range. The fields have opposite
  kinetic terms. Thus one   leads to states with positive norms and the other with negative norms.
  Which one produces positive norms and which one produces negative norms depends on the overall
  sign of the original action. We chose it in \actf\
  so that the massless field gives rise to positive norm
  states. If we had done the same computation in de-Sitter, then the sign of $m^2$  should be reversed. Note that the field $C$ transforms in a single irreducible representation of
  the four dimensional conformal group $SO(2,4)$. On the other hand, if we consider the $AdS$ problem,
  imposing suitable boundary conditions, we get two representations of $SO(2,3)$. For example,
  if we set the boundary conditions $C|_{z=0} = \partial_z C|_{z=0} =0$, then we
  get highest weight representations with $\Delta =3$ and $\Delta =2$, one representation with positive
  norms and one with negative norms. One could imagine
  changing the sign of the norm for one of these $SO(2,3)$ representations by hand, but that would be
  in conflict with four dimensional conformal symmetry.

    We note that even though
  the flat space Hamiltonian was not diagonalizable, the $AdS$ global time Hamiltonian is diagonalizable
  and has a discrete spectrum. We have the unfortunate (but expected)
  feature that many states have negative norms.
  Now, if we view $z=0$ as a boundary, it is very easy to understand why the $AdS$ global time
  Hamiltonian is diagonalizable. This ``time'' corresponds to dilatations in the plane.
    Since the equation for $C$ is conformal
  invariant we simply can consider it in flat space. Thus, dropping the denominator in \metr\ we get
  the flat space metric. The full $EAdS$ space corresponds to half of $R^4$, with a boundary at $z =0$.
  These flat space solutions can be expanded then in powers of $z$ as  $ 1, \,  z^3;~z,\,z^2$.
  The first two are the ones associated to the dimension $\Delta =3$ operator or massless field.
   The second two
   and then the one corresponding
  to the dimension $\Delta =2$ operator or tachyon field.

  It is now clear that the operator with $\Delta =2$ is sourced by the first derivative of $C$.
  In other words, the value of $C$ at $z=0$ is the source for the $\Delta =3$ operator, and $\partial_z C$
  at $z=0$ is the source for the $\Delta =2$ operator. In other words, in $EAdS$ we would set boundary
  conditions for $C= C_0(x)$ and $\partial_z C = C_0'(x)$. We can then find  a solution which decays at
  $z \to - \infty$ and obeys these boundary conditions. Since the equations are fourth order it
  is clear that we can find such a solution. These two boundary conditions simply specify the boundary
  values of the two scalars we had in \actfi .
  If we consider a Euclidean AdS space and we compute the partition function setting
   $\partial_z C (z=0) =0$ but with nonzero boundary values of $C(z=0)$,  then we excite only one of the two quadratic AdS scalars, namely
   the massless one. Thus, in this way, only the field associated with positive norms is involved.

  On the other hand, in   Lorentzian $AdS$, the $\partial_z C =0$ does not remove the ghosts.
  In fact we  have normalizable states of the tachyon field which   have negative norms. Since the
  tachyon field carries energy, it
  couples to the stress tensor. Then  a gravitational wave perturbation,
  or an insertion of the stress tensor,  can create pairs of these negative norms states\foot{
  The pair actually has positive norm. However, if we separate them by a large amount  in
  $AdS$ space we will find that each one individually has negative norm.}. We will not consider
  the Lorentzian AdS case any further.

  Let us now consider the fourth order theory in flat Minkowski space. The wavefunctional of
  this theory can be written as a function of $C$ and $\dot C$. For a quadratic theory these would
  be canonically conjugate variables. However, for a quartic theory these can be viewed as two
  ``coordinate'' variables. More precisely, we can say that the Poisson bracket between $C$ and $\dot C$ is
 zero. Thus, we can consider the wavefunctional
     $\Psi[C_B(x) , \dot C_B(x)]$. We can evaluate it at $t=0$. As usual, it can be computed either in the Lorentzian theory or in the Euclidean theory by the ordinary flat space analytic continuation.
     We can also view this as a computation of the wavefuncional in de-Sitter space,
      $ds^2 = { - dt^2 + d \vec x^2 \over t^2 }$. This is equivalent because de Sitter is
       conformal to half of
      Minkowski space, $ t \leq 0$. From the de-Sitter point of view,  we compute
     the wavefunctional in the far future, on superhorizon scales,
     because we take the $t\to 0$ limit with
     $x$ fixed.

     This wavefunctional is given by evaluating the action on a solution of the equations of
     motion with appropriate boundary conditions.
     Rather explicitly, we can write this down in Fourier space
     \eqn\solutuf{
      C(k , t) = C_B(k) ( 1  - i k t ) e^{ i k t } + C'_B(k) { t } e^{ i k t}
      }
      Inserting this into the action we get
      \eqn\classation{ \eqalign{
      i S = & - { i \over 2 } \int d^3 x \int_{-\infty}^0  dt  [( \partial_t^2  - \partial^2_x ) C]^2
      =  { i \over 2 } \int d^3 x  \left[ C \partial_t ( \partial_t^2 - \partial^2_x ) C - \partial_t C
        ( \partial_t^2 - \partial_x^2 ) C     \right]_{t=0}
       \cr
        i S = &  \int { d^3 k \over (2 \pi)^2} \left\{ - |C_B(\vec k)|^2 k^3   + | C'_B(\vec k)|^2 k   + i 2 Re[ C(\vec k) C'_B(-\vec k ) ]k^2  \right\}
       }}
       where in the first line we integrated by parts and used the equations of motion. We also used
       that for real profiles $C_B(x)$ we have $C_B(\vec k ) = C_B(-\vec k )^*$.
       These are very close to
        the Bunch Davies wavefunctions, $\Psi = e^{ i S} $,  for quadratic fields
        in de Sitter with $m^2=0$ and $m^2 =2$, with positive and negative norms
       respectively.  More precisely, if we ignore the purely imaginary term in the third line, then we
       get precisely the corresponding Bunch Davies wavefunctions. This imaginary term is a local term (involving $k^2 \sim \nabla^2 $) which does not contribute to the expectation values.
       Thus if we set $C'_B =0$ we recover the wavefunctional of an ordinary massless
       scalar field in de-Sitter.

  Then the statement is simply that this wavefunctional
  \eqn\confmra{
  \Psi_{\rm Conformal} [C_B(x),0] = \Psi_{\rm BD -quad }[C_B(x)]
  }
   where
  $\Psi_{\rm BD-quad}$ is the usual Bunch-Davies
  wavefunction evaluated in the far future, on superhorizon distances\foot{We evaluate the
  wavefunctions at fixed comoving coordinate $x$ as $t \to 0$.}, for a massless scalar with a
  quadratic action.
    The dS to EAdS analytic continuation (see section 5 in \maldaInfl ) becomes, in conformal gravity,
    the ordinary analytic continuation
    between lorentzian and Euclidean space.
    Notice that we are getting the scale invariant de Sitter wavefunction for $C_B$ from a scale invariant
    action in four dimensions. The boundary condition $\partial_t C =0$ is breaking the four
    dimensional conformal group, $SO(2,4)$, to the three dimensional one, $SO(2,3)$.

   Finally, notice that in the de Sitter context one is often interested in computing expectation
   values of observables constructed from the scalar field $C$. These can be computed by considering
   the theory on $R^4$ with the additional condition $\partial_t C =0$ at $t=0$. This looks like a
   kind of brane at $t=0$, with Neumann boundary conditions for the field $C$. Otherwise the field
   $C$ can fluctuate in an arbitrary fashion, with vacuum boundary conditions in the future and the past. We can easily compute the propagator for the field $C$
   with these boundary conditions and use it to compute expectation values.
   From the quartic conformal scalar $C$,
    with the $\dot C(t=0) =0$ boundary condition,  we get the the expectation
   value
   \eqn\propag{
    \langle C(t,\vec k) C(t',- \vec k) \rangle_{\rm conf}   \propto  { 1 \over k^3 } \left[
    (1 - i k t ) (1 + i k t') e^{ i k (t-t') }   -  t t'   k^2 ( e^{ i k (t-t')}  - e^{ i k (t+ t') } ) \right]~,
 }
 for $  t \leq t' \leq  0$ (and a similar expression for $t' \leq t$). Here we have $k = |\vec k |$.
  We can compare this with the ordinary Bunch Davies expectation values  for a massless  scalar field
  \eqn\bunchda{
    \langle \varphi(t,\vec k) \varphi(t',- \vec k) \rangle_{\rm BD-quadratic} \propto   { 1 \over k^3 }
    (1 - i k t ) (1 + i k t') e^{ i k (t-t') }  ~,~~~~~~~ t \leq t' \leq 0
    }
    we see that it agrees with the first term in \propag . In addition, they give identical results at
    $t=t'=0$, which is the main statement in \confmra . Thus, the wavefunctions are not equal for all times, they
    are only identical when they are evaluated at $t=0$. We can view the $t=0$
     slice where we impose the $\dot C =0$ condition as a ``Neumann   S-brane''.  The second term in
    \propag\ is simply due to the negative norm states which we are fixing at $t=0$.
    Note that we can analytically continue these propagators to Euclidean time if necessary.
    It is a simple matter
    to Fourier transform these propagators and express then in ordinary space. We encounter an IR
    divergence which is the usual IR divergence for a scalar field in de-Sitter space. From the quartic
    scalar we get an IR divergence also in flat space. Of course, here the two IR divergencies
     are identified with each
    other.

  Note that the simplicity of the wavefunctions for a massless scalar in de-Sitter, contained in \bunchda ,
  is ``explained'' by the connection to a conformal scalar, but with a quartic action. Recall that
  for generic masses the fixed spatial momentum wavefunctions are given by Hankel functions.

    Finally, note that expectation values can also be computed using a classical solution.
    If we consider a generating function for correlation functions $ \langle e^{\int J(x) C(x)} \rangle$,
    then its expectation value can be obtained by considering the classical solution of the Euclidean equations
    of motion with the following boundary conditions
    \eqn\equmot{
     C(x) \to 0 ~~~~{\rm  as } ~~~~ \tau \to \pm \infty ~;  ~~~~~~~~~
\partial_\tau  C|_{\tau =0} =0 ~,~~~~~~~~ { i \over 2 } \partial_\tau^3 C|_{\tau=0} =    J(x)
     }
     and $C$ continuous across $\tau =0$.

     All that we have discussed here is for a free theory, in the sense that the action was
     quadratic in the field $C$. We will now turn to the gravity case, where we will be able to
     make similar statements, but for the full non-linear theory.

\newsec{Conformal Gravity}

We now turn to the case of gravity.
The first observation is that the on shell action for four dimensional
Einstein gravity in an Einstein space that is locally
asymptotically $EAdS$ can be computed in
terms of the action of Weyl gravity \refs{\anderson,\chang,\Oleath}.
The argument is recalled in more
detail in the appendix. Here let us just give a quicker version. We can write
\eqn\tacg{
\int  W^2 =  \int e  + \int   2 ( R_{\mu \nu} R^{\mu \nu} - { 1 \over 3 } R^2 )
}
where $e = R\wedge R^* $ is the Euler density, whose integral on a closed manifold
gives a topological invariant.
Now suppose that
 we have an Einstein space, with $R_{\mu \nu} = \alpha g_{\mu \nu}$ with $\alpha$ a constant.
Evaluating the right hand side of \tacg\ we get a constant times the volume from the Ricci tensors.
Similarly the ordinary Einstein action $ \int (R - 2 \Lambda )$ gives us a constant times the volume.
Thus the on shell action is proportional to the volume, and this volume, up to a topological term, is
proportional to the Weyl action. The boundary terms that are necessary to make the bulk
integral of $e$ into
a proper topological invariant are the same as the counterterms that renormalize the volume \refs{\BKraus,\Skenderiss,\Oleaf,\anderson,\Oleas,\Oleath}, see
appendix A for more details. The final statement is that
\eqn\einconf{
 \int \sqrt{g} (R +6) - ({\rm Counterterms } )  = { 1 \over 4 } \left[  \int  \sqrt{g}\,  W^2 - E \right]  ~,~~~~~~~~E = { 32 \pi^2 \chi }
}
where $\chi$ is a topological invariant:
the Euler number of the manifold with boundary, including the boundary terms. The ``counterterms'' subtract the infinite volume of the space near the boundary.
 They are unrelated to quantum mechanical counterterms
of the bulk theory. All our computations are classical.

The second observation is that any space that is conformal to an Einstein space is a
solution to the equations of motion of conformal gravity.  The equations
 of motion of conformal gravity are  $B_{\mu \nu} =0$,  where
  $B_{\mu \nu}$ is called the ``Bach" tensor and its trace is zero because of the Weyl
symmetry of the original action, see appendix C for its explicit form.
Due to \tacg\ the Bach tensor can be expressed in terms of
derivatives of the Ricci tensor or Ricci scalar  as well as some quadratic expression in the Ricci tensor
or scalar. Let us evaluate $B_{\mu \nu}$ for an
 Einstein manifold, which obeys $R_{\mu \nu} = \alpha g_{\mu\nu}$, with
$\alpha $ a constant. Then we find that $R$ is a constant and that all the terms in $B_{\mu \nu} $ that contain
derivatives vanish automatically. All terms that do not contain derivatives can only be proportional
to $g_{\mu\nu}$, but since $B_{\mu \nu}$ is
traceless we conclude that such a term should also automatically
vanish.

Of course, the equations of motion of Weyl gravity also contain other solutions.
Here we just point out that a simple boundary condition selects the solutions that are related to
Einstein spaces.

This is argued  by spelling out the form of the well known   Starobinsky  or
 Fefferman and Graham \refs{\Starobinsky,\FG,\Skenderisf,\Skenderiss}
  expansion for a metric obeying the Einstein equations with a cosmological
 constant (an Einstein space) that is locally de-Sitter or Hyperbolic
 near the boundary
\eqn\bound{
ds^2 = {  dz^2 + dx_i dx_j \left[ g^{(0)}_{ ij}(x) + z^2 g^{(2)}_{ij}(x) + z^3 g^{(3)}_{ij}(x) + \cdots
\right]
 \over z^2 }
}
For de Sitter we have the same expansion with $ z \to i t$, up to an overall minus sign. Here we have performed the expansion in
a particular gauge (basically a  synchronous gauge) where
we have set  $g_{zz}$ to a special value  and   $g_{z i}$  to
zero.

Note that in conformal gravity we can drop the $1/z^2$ overall factor and view this as the expansion
of a manifold around $z=0$. Then the one special property of \bound\ is that there is no linear term
in $z$.

Thus, Einstein solutions are conformal to solutions which at $z=0$ obey the condition
$\partial_z g|_{z=0} =0$. We call this a ``Neumann'' boundary condition.

Given that conformal gravity has fourth order equations, we expect that it has four solutions
for a given spatial momentum.
If we are in a EAdS, or dS situation, then we require that solutions either decay or have positive
frequencies in the deep interior. This condition kills two of the solutions.
The condition $\partial_z g_{ij} =0$ kills another solution. So we are left with only one solution.
Since an Einstein space is conformal to a solution obeying all the boundary conditions,
we conclude that that is the solution that remains.

Actually the argument above was a bit too fast. Let us give a more accurate discussion that leads
to the same conclusion.
Conformal gravity has two physical modes  with helicity two and one solution with
helicity one  (and the corresponding numbers with negative helicities)\RiegertHF .
 The $\dot g_{ij} =0$ boundary
condition sets to zero one of the helicity two  modes, as well as the helicity one mode.
This can be understood explicitly by writing down the general solution of the linearized
Weyl equation (or Bach tensor).
 Writing the metric as a deformation of the flat space metric $g_{\mu \nu} = \eta_{\mu \nu} + h_{\mu \nu}$,
imposing the gauge conditions $h_{00}=h_{i0} = h_{ii} =0$ and linearizing the Bach tensor,
one can write the general solution as\foot{In this formula the indices $ij$ run over three of the dimensions.
The direction 0 could be   time or the $z$ direction, which is the direction orthogonal to the boundary.}
\eqn\gensol{
 h_{ij} \sim \left[ (  \epsilon_{  ij} + k_{(i }\zeta_{j) } )   - i k_0 t \hat \epsilon_{   ij} \right] e^{i k .x } ~,~~~~~~
 k_\mu k^\mu =0
 }
 where $\epsilon_{ij}$ and $\hat \epsilon_{ij}$ are transverse ($k_i \epsilon_{ij}=0$) and traceless ($\epsilon_{ii}=0$) and describe the two spin two
 modes.  $\zeta_i$ is also transverse ($k_i \zeta_i =0$) and it describes  the vector particle.
 We can easily see now that imposing $\dot h_{ij} =0$ at $t=0$, and a positive frequency condition in
the past,
 we set the vector mode to zero and we get only one surviving spin two mode with $\epsilon_{ij} = \hat \epsilon_{ij}$ \foot{ As a side comment, one can view this surviving mode as a massless graviton and
the vector together with the spin two mode as a ``tachyonic'' massive graviton. (It is tachyonic in the $AdS$
case, in the $dS$ case it is a massive graviton). Normally, a massive graviton also has a scalar component.
However, for this very special value of the mass, we can have a particle with only the spin two and spin one parts. This the phenomenon of ``partial masslessness'' described in \DeserPartial .
 If one added the Einstein action to the action of  conformal gravity then the
 scale factor of the metric does not decouple any longer and we get a massless and a massive graviton with generic mass (and an extra scalar component).
This type of setup was considered before with the idea of making the massive mode very massive. More recently this was also considered with the idea of making this mode degenerate with the graviton \Pope , which in effect removes the splitting that we have  between the two spin two particles in $AdS$
conformal gravity (see \actf\  for a discussion of a similar splitting between the modes of a scalar field). }. This mode then agrees with the on shell graviton mode around de Sitter. Note that
 the correspondence with conformal gravity ``explains'' the simplicity of these wavefunctions, which
 are simply plane waves up to the extra factor of time.
Conformal gravity can also be formulated as a second order theory, in a way similar to what 
we saw for the scalar field in \actfi , see appendix C of \MetsaevFQ . In that formulation one
can see more clearly the massless graviton mode, the vector and the tensor mode.

Note that the boundary condition $\partial_t g_{ij} =0$ is not invariant under Weyl transformations.
We can restrict the Weyl factor to obey $\partial_t \Omega =0$ at $t=0$. Alternatively, one could
restate all the conditions in a more general gauge independent fashion. The resulting condition
is to say that the $t=0$ slice is a ``totally geodesic'' surface \changToAppear .

The conclusion is that classically, or at the level of tree diagrams,   we have a complete equivalence
between ordinary gravity and conformal gravity. It has been observed in \Instability , that conformal gravity
around flat space suffers from a linearization instability. Namely, solutions of the linearized
equations  sometimes
do not lift to solutions of the full non-linear equations. This is due to the presence of
modes which are linear in time and the fact that the Hamiltonian cannot be diagonalized. These
features are not present if we consider the problem on a half space, or equivalently the $EAdS$ or $dS$
problems. For the case of the scalar field, we saw explicitly
 that one obtained two separated conformal
towers under $SO(2,3)$. We expect that the same happens for the graviton.
In fact, our arguments imply that conformal gravity (with the Neumann boundary condition)
 and ordinary gravity give the same answer for small but finite deformations around flat space.
 We have only checked explicitly  that  the boundary condition kills the wrong solutions at linearized
 order. However, we expect that the bulk differential operators have a spectrum with a gap, so that
 the boundary conditions continue to kill the wrong solutions in a small neighborhood of flat space.
 In particular, this is enough to establish the full tree level
 perturbative equivalence between conformal
 gravity with a Neumann boundary condition and ordinary gravity with a cosmological constant.

We conclude that at the level of tree diagrams we have the
equality
\eqn\solfg{
\Psi_{\rm Conformal }[ g, \dot g =0 ] \sim e^{  c_W   \int W^2 } = \Psi_{\rm BD - ren} ( g) =
 e^{ c_E  \int \sqrt{g} (R \pm 6 )  - (\rm counterterms ) }
}
where we have indicated that the equivalence is simply the statement that the classical actions
evaluated on the corresponding classical solutions are the same. The $ \pm$ corresponds to the $EAdS$
or $dS$ cases.
Note that conformal gravity has only a dimensionless coupling constant multiplying the whole action.
Gravity in $AdS$ or $dS$  also
  has a dimensionless coupling set by  $c \propto   (M_{pl} R)^2 $, with $M_{pl}$ the (reduced) Planck mass\foot{ $M_{pl}^2 = { 1 \over 8 \pi G_N}$.}, and $R$ the $AdS$ or $dS$ curvature radius.
   This is identified with the coupling
constant appearing in conformal gravity. More precisely the relations are
\eqn\relaconf{ \eqalign{
 c_W =&  - i { 1 \over 8 } (M_{pl} R_{dS})^2 ~,~~~~~~~~~~~~c_E = i { ( M_{pl} R_{dS})^2 \over 2 } =
 i { M^2_{pl} \over 2 H^2}
~,~~~~{\rm for ~ de~Sitter}
\cr
c_W =&   { 1 \over 8 } (M_{pl} R_{EAdS})^2 ~,~~~~~~~~~~~~c_E = { ( M_{pl} R_{EAdS})^2 \over 2 }
~,~~~~{\rm for ~ Euclidean~AdS  }
}}
These expressions could also be written in terms of the cosmological constant, which we have set here
to $ \Lambda = \pm { 3 \over R^2 }$.
Note that in Euclidean Anti de Sitter we get the Weyl action with the ``wrong'' sign. Namely, the
Weyl action has the nice feature that it is bounded below in Euclidean space. However, in the
Euclidean Anti-de Sitter context we get it with the opposite sign. This is no problem in perturbation
theory. Furthermore, this is physically reasonable since the two point function of the stress tensor,
given by ${ \delta \over \delta  g^{ij}(x) } { \delta  \over \delta g^{kl}(y) } \Psi[ g] $, should
be positive if the Euclidean gravity theory corresponds to the Euclidean continuation of a unitary
boundary CFT (which is usually the case in $AdS/CFT$).
On the other hand, the sign we get in the de-Sitter case is such that if we do the usual analytic
continuation to Euclidean signature we get the ``right'' sign of the $W^2$ action.
These facts are, of course, consistent with the observation in \refs{\maldaInfl,\Harlow} that
 the de Sitter wavefunction can be obtained from the EAdS one by simply
flipping the sign of $(M_{pl} R)^2$. This is just an overall sign in the exponent.

This connection might
lead to a  practical way to evaluate tree diagrams in $AdS$ or $dS$, since in
both cases we could view the computation as a computation around flat space in conformal gravity.
The propagators and vertices of the two actions are different, but they both should give the
same final answer. Given that twistor string theory contains conformal gravity \BerkovitsWitten ,
 maybe one can
also use that string theory to compute de Sitter correlators.

\ifig\diagrams{ (a) An example of a tree Feynman-Witten  diagram that contributes to the
 tree level computation
of the AdS partition function or the de Sitter wavefunction. We are stating that these diagrams
give the same answer regardless of whether we compute them in Einstein gravity or in conformal
gravity.  (b) An example of a loop diagram that
is {\it not} contained in the discussions of the present paper. (c) One example of a tree diagram
that contributes to expectation values in de-Sitter. The top and bottom denote the two branches
of the Schwinger-Keldysh contour.  Alternatively, they can denote the upper or lower half space after
analytic continuation to Euclidean signature.
  } {\epsfxsize2.5in\epsfbox{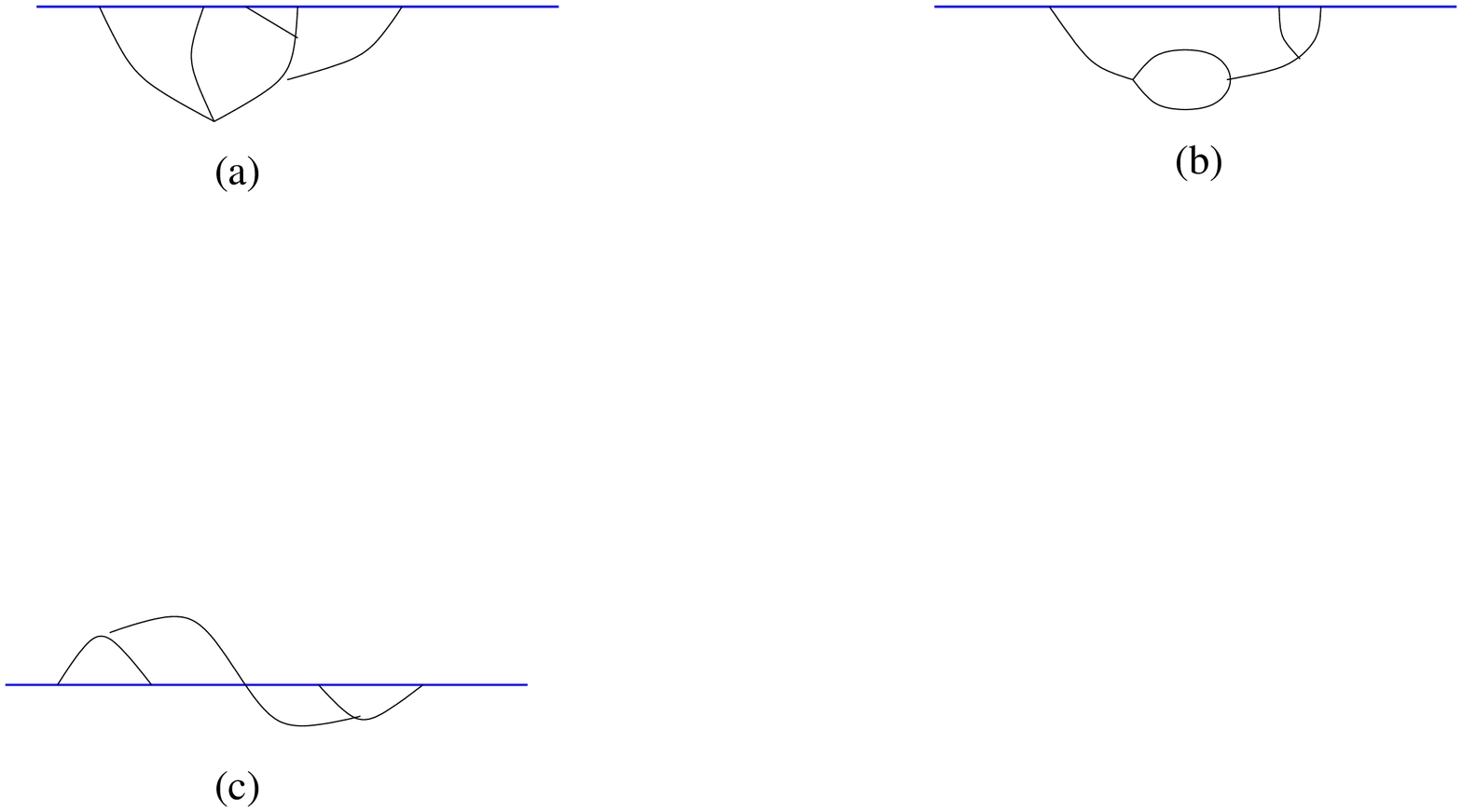}}

In the case of $dS$ computations, it is natural to integrate over the metric on the boundary and
compute $\int {\cal D} g |\Psi(g)|^2$ \foot{ Of course, this integral should be defined carefully.
In particular, we do not integrate over the overall scale factor in the metric. In conformal gravity,
this is clear, since that is gauge symmetry. In the de Sitter case, this can be viewed as
 selecting a ``time'' slice. Also, at tree level we just use saddle point and
 the details of the measure  do not matter. }.
  In conformal gravity this can be viewed as computing expectation
values for the metric or the Weyl tensor in the presence of a
   Neumann S-brane, which is defined by setting $\partial_z g_{ij} =0$. Namely, we set $\partial_z g_{ij} |_{z=0}=0$
but we integrate over the value of $g_{ij}$ at $z=0$. At tree level this can be done, as in the case
of the scalar field, by looking for classical solution of the Euclidean equations of motion with the
boundary conditions that the space becomes flat as $\tau \to \pm \infty$ and that
the metric is continuous at $\tau =0$, and it obeys
\eqn\metrico{
\partial_\tau g_{ij}|_{\tau =0}  =0 ~,~~~~~~~~ i \partial_\tau^3 g_{jl}|_{\tau =0}  = J_{jl}(\vec x )
}
Then the classical action evaluated on this solution gives us
$e^{-c S_{clas}} = \langle e^{ \int J_{ij} h_{jl} } \rangle $, here $h$ is a deformation from flat space,
$g_{\mu \nu} = \eta_{\mu \nu} + h_{\mu \nu}$.

\newsec{Black holes}

The arguments we presented above establish the equivalence for small (but finite) deformations around
flat space (or an $S^3$) on the boundary.
 We could wonder if there are new solutions  once we consider  large enough deformations.
As an initial exploration, we consider a boundary of the form $S^1 \times S^2$, so
that we get contributions from
Euclidean black holes. Black hole solutions in conformal gravity were considered in
\refs{\Bach,\RiegertZZ,\DeserUP}.
We can write down the general spherically symmetric ansatz for conformal gravity as
\eqn\genans{ds^2 = e^{2 g}  dt^2 + { d\rho^2 \over e^{2 g } }  + d\Omega_2^2
}
where $d\Omega_2$ is the metric of the two sphere. We have used the fact that, in conformal gravity,
 we can perform
a Weyl  transformation that sets the  radius of the two sphere to one.
The equations of motion of conformal gravity imply
\eqn\equbl{
-(g'') ^2+4 (g' )^4+2 g^{(3)}  g' +8 (g') ^2 g'' +e^{-4
   g }=0
   }
Note that the equation is cubic and not quartic. This reduction in order also arises in Einstein
gravity and it is due to the reparametrization symmetry constraint\foot{In other words,
if we   introduce  $N$ via   $d \rho^2 \to N d \rho^2 $ in \genans , then \equbl\ is the
equation of motion for $N$, at $N=1$.}. We set the boundary at $\rho=0$, and set $g'(0)=0$ at this
point. We can also use
a symmetry under rescalings of the coordinates to set $ g(\rho =0) =0$.
  Since \equbl\ is a cubic equation we expect a three parameter family of solutions. We have
  already fixed two of the parameters by using symmetries of the equation. Thus the solution
  depends only on one non-trivial parameter.
The general solution is
\eqn\anstaso{
e^{ 2 g } = (1 \pm  \rho^2  - 2 m  \rho ^3 )
}
There are two branches of solutions, associated to the  $\pm$ signs.
 Let us discuss first the branch with the plus sign, or $g''(0)=1$.
This branch describes the ordinary $AdS$ black holes, with $m > 0$. Of course, $m=0$ describes
global $AdS$.
 In other words, up to an overall
 factor of $r^2$, the
metric \genans\ is the same as \HP
\eqn\ordsol{
ds^2 =  ( r^2 + 1 - 2 m/r ) dt^2 +  { dr^2 \over  ( r^2 + 1 - 2 m/r ) } + r^2 d\Omega^2_2
}
together with the identification $\rho = 1/r$.
The free energy of the black hole can be computed using conformal gravity. We get
\eqn\usew{  { M_{pl}^2 R_{AdS}^2 \over 8 } \int W^2 = 8 \pi^2 M_{pl}^2 R_{AdS}^2
 \left[ { (1 + \rho_h^2)^2 \over \rho_h^2 (3 + \rho_h^2 ) } \right]  ~;~~~~~~~~~ \beta = { 4 \pi \rho_h
 \over     {  \rho_h^2  } + 3  }
}
where $\rho_h$ is a root of $e^{2g} =0$.
In particular, the
entropy of the solution can also be computed using Wald's formula  \WaldNT .
The Wald entropy of a black hole in a theory where the lagrangian depends on the curvature, $L(g,R_{\mu\nu\rho\sigma})$,
is given by \JacobsonVJ\
\eqn\waldent{
S = - 2 \pi \int_{\Sigma_2} L^{\mu \nu \rho \sigma} \epsilon_{\mu \nu} \epsilon_{\rho \sigma} ~,~~~~~~~~
 L^{\mu \nu \rho \sigma} = { \partial L \over
 \partial R_ {\mu \nu \rho \sigma} } \propto W^{\mu\nu \rho \sigma }
}
where $\Sigma_2$ is the horizon and $\epsilon_{\mu \nu}$ is the binormal, normalized
so that $\epsilon_{\mu \nu} \epsilon^{\mu \nu} = -2$.
For the above solution this simply gives
\eqn\giveso{
S = {  4 \pi R_{AdS}^2 (r_h^2 + 1 ) \over 4 G_N }
}
where $r_h$ is the horizon radius.
The constant factor is due to the contribution from the Euler character. Since the black hole has
a different topology, the Euler character  in \einconf\  gives us a non-trivial contribution.
(The Euler character of the black hole is $\chi =2$). This extra contribution
to the partition function (or $\beta F$)
is independent of the temperature and thus, it gives a constant contribution to the entropy\foot{
As a completely side remark, notice that if we have Einstein gravity plus a correction proportional
to the Euler character, then the fact that the black hole entropy should be positive sets a
bound on the size of the coefficient of the Euler character term. This puts a bound for one sign
of the coefficient. For the other sign of the coefficient, we can set a bound by
demanding that the entropy of the black hole is less than that of the Hawking radiation once it has
evaporated completely (this bound depends on the number of light species). The  bound is set by the
lowest distance scale at which we trust the black hole solutions. The bound on the Euler coefficient is
 then  of the order of $ (4 \pi)^2 { M^2_{pl} \over M_s^2 }$ where $M_s$ is the energy scale at which we cease to  trust Einstein gravity.  }.
As usual, these Schwarschild AdS black holes exist for inverse
temperatures bigger than  $\beta \leq { 2 \pi \over \sqrt{3}} $.

One can ask the following question. Imagine that we evaluate the Wald entropy formula \waldent\ for
conformal gravity, on a solution of Einstein's gravity. Then can we show that it is equal to the area?.
In fact, it is easy to show that this is the case. Writing  $\int W^2 - E$ as in \tacg\ , the Wald's
formula gives us terms involving the Ricci tensor. If we have an Einstein space these become simply
the metric. Thus, it is always automatic that Wald's formula on conformal gravity (minus the Euler
character) on an Einstein space reproduces the results of Einstein gravity. Of course, this is
also a consequence of the equivalence between the two on shell actions.

In addition to the ordinary black hole solutions,  the equations of conformal gravity have
a second branch of solutions if we choose the minus sign in \anstaso , or $g''(0)=-1$.
These are additional solutions which exist with our boundary conditions. There is a single solution
for each value of the temperature.
The ``origin'' of these solutions can be understood as follows. In Einstein gravity,
we have a related family
of black hole solutions obtained by replacing  $S^2$ by $H^2$
and $(r^2 + 1 - 2 m/r ) \to (r^2 -1 - 2 m/r)$ in
\ordsol\   \Emparan , with $m\geq -{ 1 \over \sqrt{27}}$.
In conformal gravity these are also solutions of the $S^2$ problem for the following reason.
Starting from $H^2$ we can make an analytic continuation which takes $ds^2_{H^2} \to - ds^2_{S^2}$.
We can also take $e^{2g} \to - e^{2g }$, which is a symmetry of \equbl . Finally we can make
a Weyl transformation by an overall minus sign in the metric. This chain of arguments explains
``why'' we found another solution for the $S^2$ problem. Of course, it is a simple matter to check
that \anstaso\ is a solution of \equbl . The two branches are distinguished by the value of
$g''(0)$ and we can imagine selecting the correct branch by selecting the appropriate sign.
In fact, in an Einstein space, the Fefferman Graham expansion \bound\  fixes the value of the second
derivative of the metric around the boundary. However, from the point of view of conformal gravity
this seems unnatural.

\newsec{ The Hartle Hawking factor and the sphere partition functions }

As a side remark, notice that the de Sitter
 Hartle Hawking factor can also be obtained in all dimensions
from the analytic continuation from Euclidean AdS, see also \Harlow .
 In the case of even bulk dimensions, this Hartle
Hawking factor is the same (more precisely it is the square) of the sphere partition function of the
boundary field theory (if that theory were to exist). Namely, it is the de Sitter analog of the
sphere partition functions that have been recently   computed in \refs{\Kapustin,\Marino,\Jafferis}, and
references therein. For example, consider the case of four bulk dimensions, and three boundary dimensions. If we have an $S^3$ boundary the field theory partition function for a theory that has an $AdS_4$ dual
is computed by
\eqn\comput{\eqalign{
 \log Z= &   { ( M_{pl} R_{AdS} )^{2   } \over 2 } v_{S^3}  (- 6) \int_0^{\rho_c} d\rho (\sinh \rho)^3 ; \cr
& ~~~ \int_0^{\rho_c} d\rho (\sinh \rho)^3 =   { e^{ 3 \rho_c} \over 24 } - { 3 e^{ \rho_c} \over 8 } + { 2 \over 3 }  + o ( e^{-\rho_c} )
}}
The sphere partition function is obtained by taking the $\rho_c \to \infty $ limit, discarding the
divergent terms. Thus, it is given by the factor of $2/3$ in the square brackets. Here $v_{S^3} = 2 \pi^2$
is the volume of a three sphere.
On the other hand, we can compute the Hartle Hawking factor
\eqn\hhfa{\eqalign{
 \log |\Psi|^2_{HH}   = &
 { ( M_{pl} R_{dS} )^{2  } \over 2 } v_{S^{3} } 6 \left[  2 \int_0^{\pi \over 2 }
 d\rho (\sin \rho)^{3 } \right]
\cr & ~~\int_0^{\pi \over 2 }
 d\rho (\sin \rho)^{3 }  = { 2 \over 3 }
 }}

These two factors are equal (up to a factor of two , since we need to multiply  \comput\
by to to take into account that we have $|\Psi|^2$ in
 \hhfa ). The equality holds, after we analytically continue $R_{AdS}^2 \to - R_{dS}^2 $.
We can formally go from \comput\ to \hhfa \ by taking $\rho \to \rho + i \pi/2$ and $R_{AdS} \to - i R_{dS}$. See a more detailed discussion in \Harlow .

Thus, from a dS/CFT perspective, the Hartle Hawking factor is the $S^3$ partition function of
the dual field theory, with the local infinities subtracted.

Of course, if one were to replace the $S^3$ by other manifolds, such as a $S^1 \times S^2$, or $T^3$,
then one would get different answers. In fact, black branes in $AdS_4$ can be viewed as computing such
factors, due to the analytic continuation from $EAdS$ to $dS$.

This works in a similar way in other even bulk dimensions. For odd bulk dimensions, there is a
logarithmic divergence and the Hartle Hawking factor is related to the coefficient of this
divergence, as we will see explicitly for the five dimensional case  in the next section.

\newsec{ Four dimensional conformal gravity and five dimensional Einstein gravity}

In this section we discuss a different appearance of four dimensional conformal gravity from de Sitter
space. We will show that 4 dimensional Euclidean conformal gravity is the late time (superhorizon)
measure factor arising from tree level five dimensional gravity in de-Sitter space.

Before considering the de Sitter problem, let us review a well known property of the
Anti-de-Sitter case.
Let us consider evaluating the action for an asymptotically hyperbolic space which is a
Euclidean solution of five dimensional Einstein gravity with a cosmological constant.
We fix a boundary metric of the form $  { \hat g_{ij} \over \epsilon^2 } $. We also have a
bulk Einstein metric, obeying the usual five dimensional Einstein manifold condition, $R_{\mu \nu} = - 4 g_{\mu \nu} $ with a boundary condition set by the four dimensional metric $\hat g$.

Then  the on shell Euclidean action ($ Z \sim e^{ - S} $)
 has the following expansion \Henningson
\eqn\estruct{\eqalign{
 -  S(\hat g) = &
{M_{pl}^3 R_{AdS}^3 \over 2 } \left[
  \int d^5 x  \sqrt{g} ( R + 12 )  + 2 \int d^4 x K  \right] =
  \cr = &  {M_{pl}^3 R_{AdS}^3 \over 2 } \left[ { a_0  \over \epsilon^4 } \int d^4 x  \sqrt{\hat g } +
   { a_2  \over \epsilon^2 } \int d^4 x \sqrt{\hat g} \hat R  +
{  \log \epsilon \over 8} \int \sqrt{\hat g}    ( \hat  W^2 - \hat e )  \right] - S_R[ \hat g ]
}}
where $S_R$ is finite as  $\epsilon \to 0$ and $a_0$, $a_2$ are two (real) numerical constants\foot{$a_0 =6$. In $d$ boundary dimensions, this first coefficient is $a_0 = 2 (d-1)$, which is positive.}.

We recognize the action of conformal gravity in the coefficient of the logarithmic term \refs{\LiuBU}
\foot{If we had maximally supersymmetric 5d gauge supergravity in the bulk, then we would get
the action of ${\cal N}=4$ conformal supergravity as the coefficient of the logarithmic term \refs{\LiuBU}.}.

Now, let us consider the case of pure de-Sitter gravity in five dimensions.
In that case, we can similarly evaluate the wavefunction, as we did in four dimensions.
This wavefunction can be evaluated from analytic continuation from the Euclidean $AdS$ case
we mentioned above. Namely, we set
\eqn\changes{ z = - i \eta ~,~~~~~~~ R_{AdS} = - i R_{dS}
 }
This implies that the cutoff $\epsilon_z = - i \epsilon_{\eta}$. In de Sitter, $\eta, ~\epsilon_{\eta} < 0 $.
As in the four dimensional case, all divergent ``counterterms'' become imaginary, so that that they
drop out from the quantum measure, $|\Psi|^2 $. On the other hand, in five dimensions,
the finite part, given by $S_R$,
{\it also} becomes purely imaginary and drops out.
The only real term comes from the analytic continuation of the logarithmic term
$\log \epsilon_z \to \log (-\epsilon_\eta )  + i \pi/2 $.
This gives a square of the wavefunction of the form
\eqn\squarewf{
 \left| \Psi[ \hat g]  \right|^2_{\rm Einstein }  = \exp \left[  -  {M_{pl}^3 R_{dS}^3 \over 2 }    { \pi \over 8 } \int d^4 x \sqrt{g}  ( \hat W^2 - \hat e )  \right]
 }
 One can check explicitly that this gives the right value for the quadratic fluctuations \maldaInfl .
 Here we claim that this captures all the tree level diagrams in de-Sitter space.
 One remarkable feature of this action is that it is purely local in space. This is in contrast to the
 four dimensional case, where $|\Psi|^2 $ has a non-local expression.
 This is, of course, just a tree level result. As we include loops we could generate non-local terms
 in the effective action. Loop diagrams would not have the factor of $(MR)^3$ which will make their
 analytic continuation to de-Sitter different.

 Now, if we want to compute expectation values of the metric in five dimensional de-Sitter space, then
 we could do it in the following way. First we compute $|\Psi[ \hat g] |^2$, which gives the action of
 four dimensional conformal gravity.
 Then we can do the functional integration over boundary metrics. This then becomes a problem in
 four dimensional conformal gravity. Again, if we restrict to tree level results, we have a
 precise equivalence with tree level computations in four dimensional conformal gravity.
 Note that in the five dimensional de-Sitter problem we can compute probabilities by slicing the geometry
 at various times. In terms of the wavefunction of the universe, this translates into various scale
 factors for the geometry. The measure $|\Psi|^2\sim e^{ - S_{\rm conf} } $
 is explicitly independent under the choice of scale factor for the spatial metric due to the exact
  Weyl invariance of the action of four dimensional gravity. Furthermore, when we compute expectation
  values with this measure we { \it do not } integrate over the scale factor. From the point of
  view of the de-Sitter, we do not integrate because probabilities from the Wheeler de Witt wavefunction
  are computed at a given time (or scale factor). From the point of view of conformal gravity, the
  invariance under rescalings of the scale factor is a gauge symmetry, thus, we ``divide by the volume of
  the gauge group'', which amounts to fixing a particular scale factor.

 In summary, {\it late times, superhorizon,
 expectation values of the metric in five dimensional de-Sitter space are equal, at tree level,
 to expectation values computed using four dimensional conformal gravity }. In other words,
 the tree level cosmological superhorizon measure in  five dimensional gravity with a cosmological
 constant   is given by four dimensional conformal gravity.

 As a simple check of these formulas, let us compute the five dimensional Hartle Hawking factor using
\squarewf . The Hartle-Hawking factor is a real term that can be interpreted as giving us the
probability of making a universe with $S^4$ spatial topology \HH .
 (Recall that we are in five dimensions, so
 $S^4$ is the spatial slice.) It is given by evaluating the five
dimensional Einstein action on an $S^5$ (and it equals the entropy of $dS_5$). It is
\eqn\hhfa{
|\Psi|^2 \sim  \exp \left[  { M_{pl}^3 R_{dS}^3 \over 2 } \int_{S^5} \sqrt{g} ( R - 12 )  \right] =
  \exp \left[{ M_{pl}^3 R_{dS}^3 \over 2 } 8 \pi^3  \right]
}
where $\pi^3$ is the volume of the five sphere.

We get the same from \squarewf. For $S^4$ the Weyl tensor vanishes because $S^4$ is conformally flat.
Then only the piece involving the Euler number survives, where $\chi(S^4) =2$.
Thus we get
\eqn\geteu{
|\Psi|^2 \sim \exp \left[   { M_{pl}^3 R_{dS}^3 \over 2 } \pi  4 \pi^2 \chi(S^4) \right] = \exp \left[  { M_{pl}^3 R_{dS}^3 \over 2 } 8 \pi^3 \right]
}

 \newsec{Discussion}

We have shown that four dimensional
conformal gravity with a Neumann boundary condition is classically equivalent
to ordinary four dimensional Einstein gravity with a cosmological constant.
 This equivalence was shown only for small but finite deformations around a flat
 (or $S^3$ boundaries). These two theories are equivalent for the computation of the
 ``renormalized'' wavefunction, or partition function. In the $EAdS$ context conformal
 gravity computes the renormalized partition function. In the $dS$ case, it computes the
 superhorizon part of the wavefunction. Alternatively, we can say that it computes
 the tree diagrams for gravity wave fluctuations outside the horizon, see \diagrams .
 This relation might be useful for computing tree diagrams in de Sitter or AdS.
 For large deformations of the   geometry we found new solutions, though it might be
 the case that one can get rid of these extra solutions. We considered boundaries with
 $S^1 \times S^2$ geometry which give rise to bulk configurations corresponding to
  Euclidean black holes.

 It would be  more exciting if we could also use conformal gravity to
 compute such wavefunctions at the quantum level. The reason is that a  ${\cal N}=4$
 supersymmetric version is believed to be finite \foot{ Loop corrections in pure bosonic
   conformal gravity   generate bulk logarithmic divergencies.
    While in quantum field theory such divergencies
   are not a problem (if the theory is asymptotically free),
    in the context of conformal gravity, where the Weyl symmetry is a gauge symmetry, we
should discard these theories as anomalous   \StromingerConformal . These divergencies cancel in  some specific versions of ${\cal N}=4$ conformal gravity, one
  of which involves coupling it to   a rank four Super Yang Mills theory \refs{\FradkinBeta,\Fradkin,\Axial,\BerkovitsWitten,\Fradkin }.
}.
  One could compute the wavefunctions at one loop in conformal gravity with the
  Neumann boundary condition. These are interpreted as
  superhorizon wavefunctions and it is not clear
  how to check that they correspond to wavefunctions in a unitary theory.
    Maybe the quantum corrected wavefunction is not Weyl  invariant.

Another interesting   problem is to generalize the Neumann boundary
condition to the full ${\cal N}=4$ conformal gravity case.   Here one would expect to
put boundary conditions that remove some of the fields of the conformal gravity theory.
A quick look at Table 1 of \BerkovitsWitten , shows that if we remove all fields which
do not come in doublets, and we retain one field per doublet, then we get the physical field content
of $SO(4)$ gauged supergravity in four dimensions\foot{This was suggested by N. Berkovits.}. However, we have not found the full non-linear embedding of a solution of SO(4) gauged supergravity into ${\cal N}=4$ conformal supergravity. It would be interesting
 whether this work (or some variation of it).

${\cal N}=4$ conformal gravity also includes a scalar field with dimension zero, like the one we discussed in section two.
In conformal gravity it is  possible to couple the scalar field to the Weyl tensor as
$\int f(C) W^2 + \cdots $. This function cannot be removed by a Weyl transformation, as in Einstein
gravity. More precisely, we have a complex field $C$ such that
the action has the form $\int f(C) W_+^2 + f(C)^* W_-^2 + \cdots $.
The conformal gravity theory that arises from the twistor string theory has the form
$S = \int e^C W_+^2 + \cdots $.
Now, as we mentioned  around \relaconf\  this overall factor in the action is the ratio of the Planck scale to the
de-Sitter radius. Thus, in such a theory, one could get a very large ratio between these two scales
with a moderate value of $C$.
The theory is   reminiscent of Liouville theory in two dimensions.

We have also mentioned the fact that the de Sitter Hartle Hawking factor, related to the
de-Sitter entropy, $e^{ - S_{dS}}$ can be viewed as the $S^3$ partition function of a
hypothetical dual field theory that lives at the $S^3$ boundary of $dS_4$. This is a side
remark, independent of the relation with conformal gravity.

This could be generalized to higher even dimensions.
In six dimensions, we need two boundary conditions on the metric, on the first derivative and on  the
second derivative. In that case the bulk action is sixth order and it has the rough form $\int W \nabla^2 W $ (the precise for is given en eqn. (3.6)  of \chang ).

It would be interesting to understand whether the flat space tree level S-matrix of pure gravity
could also be computed in terms of a computation in conformal gravity. It seems that if we compute
amplitudes in conformal gravity which involve only the spin two
solutions that do not increase in time, then
we could   select the Ricci flat solutions. The amplitude should then follow from
evaluating a suitable boundary term, since the on shell bulk action would be zero.

Going now to five dimensions, we have shown that the square of the
wavefunction of the universe of five
dimensional pure gravity is given by the action of conformal gravity.
This follows in a simple way from the well known expansion of the wavefunction at late times and
the connection (via analytic continuation) to the Euclidean $AdS_5$ problem.
It seems surprising that the wavefunction is completely local, in stark contrast with the
 one in $dS_4$.

{\bf Acknowledgments}

I am specially grateful to Alice Chang for bringing \refs{\anderson,\chang} to my attention.
I also thank N. Arkani Hamed,
N. Berkovits, S. Deser, D. Harlow, D. Hofman, A. Strominger, A. Tseytlin and E. Witten for discussions.
This work was supported in part by  U.S.~Department of Energy grant \#DE-FG02-90ER40542.

\appendix{A}{ Renormalized Einstein action as the conformal gravity action}

 Anderson showed that  renormalized volumes in asymptotically hyperbolic Einstein
 spaces are given by the Weyl gravity action \anderson .
 Here we simply outline his argument.

 We are interested in computing the renormalized Einstein action, given by \refs{\BKraus,\Skenderiss}
 \eqn\renact{
 -S_{E,ren} =  \lim_{\epsilon \to 0 }  \left[
 \int_{\Sigma_{4,\epsilon} }  ( R + 6 ) + 2 K   -  4 \int_{\Sigma_3 }  \sqrt{g} - \int_{\Sigma_3} \sqrt{g} R^{(3)} \right]
 }
 where we have set the $AdS$ radius to one.
 For an Einstein manifold we have that $R_{\mu \nu} = - 3 g_{\mu \nu}$, so that evaluating the
 Einstein action is the same (up to a coefficient) to computing the volume.
  More precisely, we get
  \eqn\renact{
 -S_{E,ren} = - 6 V_{ren} = - 6  \lim_{\epsilon \to 0 }  \left[
 \int_{\Sigma_{4 } }  \sqrt{g}    -  { 1 \over 3 } \int_{\Sigma_3 }  \sqrt{g} - { 1 \over 2 } \int_{\Sigma_3} \sqrt{g} R^{(3)} \right]
 }
 In order to evaluate this \anderson\ starts from the expression for the Euler density
  $e = R_{\alpha \beta \gamma \delta} R^{\alpha \beta \gamma \delta} - 4
R_{\alpha \beta  } R^{\alpha \beta  } +   R^2$, so that $E = \int \sqrt{g} e$ is a topological invariant
on a closed manifold.
 We then use the fact that
 \eqn\farc{
   W^2 - e = 2 ( R_{\mu \nu} R^{\mu \nu } - { 1 \over 3 } R^2 )  ~~~~\to ~~~ - 6 . 4
  }
  where we have written the form of the right hand side for an Einstein space.
  Thus we see that integrating the left hand side  we get, up to a number, the volume of the space.
  At this point we might be confused because the right hand side is infinite, while the integral of
  $W^2$ would be finite. What happens is that the integral of $e$ is not a topological invariant unless we add some boundary terms. These boundary terms have a simple Chern Simons form, so that the
  full topological invariant is
  \eqn\topoin{
  E =   \int_M  \epsilon_{abcd} R^{ab} \wedge  R^{cd} -
\int_{\partial M }  \epsilon^{abcd}
 [ \omega^{ab} d\omega^{cd} + { 2 \over 3 } \omega^{ab} (\omega^{cf} \omega^{f d} ) ]
}
where $w$ is the spin connection.
This differs from what we had in \farc\ by this last term. This last term is actually divergent
as we approach the $AdS$ boundary. In fact, for an Einstein space, it gives purely divergent terms
which are the counterterms that renormalize the Einstein term and no finite residual term.
This is expected since the divergent terms have to cancel out the divergences properly and there
is no local invariant expression of the boundary metric that could remain as a possible finite term.
Thus in conclusion we find that
  \eqn\farcn{
  { 1 \over 4 } \left[ \int W^2 - E \right]  =    S_{Einstein-ren}  ~,~~~~~~~~\chi = { 1 \over 32 \pi^2 } E
  }
 Note that the left hand side is now fully conformal invariant and it can be computed with any
 metric in the same conformal class. In particular, by dropping the $1/z^2$ factor in \bound\  we get a
 metric which is a small deviation around flat space. Moreover, if we consider small deviations from
 $AdS$, then we find that $E=0$ (since it is zero for flat space and it is a topological invariant).
 The conventionally normalized Euler character is $\chi$ in \farcn.

 This argument has been extended to other dimensions in \chang .

\appendix{B}{ Norms of states in conformal gravity}

In a conformal field theory it is useful to consider the theory on $S^3 \times R$ and to
study the states of the theory on such a space.
We can do the same for conformal gravity and we can classify the quadratic fluctuations
using the conformal symmetry. In other words, we fix the background metric to $S^3 \times R$
and we study the linearized theory, with a linearized gauge symmetry.
In the full theory, the conformal symmetries are constraints on the quantum state of the theory,
so that a full non-linear analysis is more complicated and it will involve some evolution on the
background metric. We ignore this in our discussion. We also use the usual state-operator mapping
to discuss the states of the theory.

We simply would like to point out that all the states of a graviton are in a single representation
of the conformal group. Via the state operator correspondence, we find that all states are
in a single representation of the conformal group. They are all descendents of the
Weyl tensor $W$. More precisely, we have two representations, one coming from $W^+$ and one from $W^-$.

It is interesting to compute the norms of these states. All norms can be computed by the use of
the conformal algebra commutation relations, $[K,P] \sim D + M $ and $P^\dagger = K$  relation.
All norms are then computed in terms of the norm of the lowest weight state. Since this state
is simply the Weyl operator, it has conformal dimension two and spin two. $(\Delta, S) = (2,2)$ lies
outside the
unitarity bound and for this reason its descendents   have both positive norm and negative norm
states.
This computation also shows that there is a null state (orthogonal to all states) which is
simply zero by the (linearized) equations of motion for the Weyl tensor.

It is also interesting to consider the scalar field $C$. Naively we should start from a conformal
dimension zero primary. However, as in two dimensions, it is better to view $\partial_\mu C(0)$ as
the primary. Its descendents then have both positive and negative norms. There is a null state, which
is simply the equation of motion for $C$. The field $C$ itself is not a good conformal field, due
to IR divergencies. Exponentials such as $e^{\alpha C}$ are good conformal operators.
The field $C$ is similar to the scalar  field of Liouville theory in two dimensions.

\appendix{C}{Formulas for the Bach tensor}

Here we write a few useful formulas.
 We follow Wald's conventions \Wald . The following divergence of the Weyl tensor is
\eqn\equat{
\nabla_\alpha W^\alpha_{~ \mu\nu \rho } = - { 1 \over 2 } \left[
\nabla_\rho R_{\mu\nu} - \nabla_\nu R_{\mu \rho} + { 1 \over 6 } ( g_{\mu\rho} \nabla_\nu R -
g_{\mu\nu} \nabla_\rho R ) \right]  = C_{\mu\nu\rho}
}
We have that $\nabla^\mu C_{\mu\nu\rho} =0$, $C_{[\mu\nu\rho]} =0$, $C_{\mu\nu\rho}$
is called the Cotton tensor.
The equations of motion of conformal gravity involve the Bach tensor \Bach\
\eqn\bachte{ \eqalign{
B_{\mu\nu} = &  2 \nabla^\beta\nabla^\alpha W_{\alpha \mu \nu\beta} +  W_{\alpha \mu\nu\beta} R^{\alpha
\beta} =
 2 \nabla^\alpha\nabla^\beta W_{\alpha \mu \nu\beta} +   W_{\alpha \mu\nu\beta} R^{\alpha
\beta}
 \cr
 B_{\mu \nu} =   &  \nabla_{\alpha} \nabla_\mu   R^{\alpha}_\nu + \nabla_{\alpha} \nabla_\nu R^{\alpha}_\mu  -
\nabla^2   R_{\mu \nu }  - { 2 \over 3 } \nabla_\mu \nabla_\nu R + { 1 \over 6 } g_{\mu \nu } \nabla^2 R
\cr
& - 2 R_{\alpha \mu } R^{\alpha}_{\, \nu}  + { 2 \over  3} R^{\mu \nu} R  + { 1 \over 2 } g_{\mu \nu}( R_{\alpha \beta} R^{\alpha \beta} - { 1 \over 3 } R^2 )
 }}

\listrefs

\bye